\documentclass[a4paper,twoside]{article}
\usepackage{microtype}
\usepackage{wrapfig}
\usepackage{epsfig}
\usepackage{subcaption}
\usepackage{calc}
\usepackage{amssymb}
\usepackage{amstext}
\usepackage{amsmath}
\usepackage{amsthm}
\usepackage{multicol}
\usepackage{pslatex}
\usepackage{apalike}
\usepackage{algorithm2e}
\usepackage[bottom]{footmisc}
\usepackage{xcolor}
\usepackage{enumitem}
\usepackage{url}
\setlist{nosep}
\usepackage{SCITEPRESS} 

\begin{document}

%% --- Math Squeeze ---
%\setlength{\abovedisplayskip}{4pt}
%\setlength{\belowdisplayskip}{4pt}
%\setlength{\abovedisplayshortskip}{0pt}
%\setlength{\belowdisplayshortskip}{0pt}

\title{MIMO and Multi-Group Comparison Problem in Process Control: A Multivariate Statistical Framework for Systems Represented with Frequency Response Functions}

\author{\authorname{Vittorio Lippi\sup{1}\orcidAuthor{0000-0001-5520-8974}, Leonard Johard\sup{1}\orcidAuthor{0000-0002-8222-9815} and Gabriele Landucci\sup{2}\orcidAuthor{0000-0002-0212-3253}}
\affiliation{\sup{1}Calejo Hybrid Intelligence AB, Mellansj\"{o} 109, 92592 T\"{a}rnaby, Sweden}
\affiliation{\sup{2}Department of Civil and Industrial Engineering Largo Lucio Lazzarino, 1, 56126, Pisa, Italy}
\email{\{vittorio.lippi, leonard.johard\}@calejo.ai, gabriele.landucci@unipi.it}
}


\abstract{Characterizing complex Multi-Input Multi-Output (MIMO) systems presents two issues: feedback controllers mask fault variance, rendering single-variable monitoring ineffective, and repeated testing across multiple conditions inflates false positive rates due to the multiple comparisons problem. This study proposes a multivariate statistical framework to resolve these limitations. We extend a statistical library that identifies system dynamics via \textit{Frequency Response Functions} (FRFs) by transforming them into time-domain \textit{Pseudo-Impulse Responses} (PIRs). This functional representation captures the complete dynamic signature of the system, offering a richer diagnostic profile than traditional static scalar metrics. The framework, originally developed for SISO systems, is extended to the MIMO case by introducing supervectors that aggregate multiple PIRs, evaluated using \textit{Permutational Multivariate Analysis of Variance} (PERMANOVA). This non-parametric approach handles the high dimensionality and complex correlation structures inherent in MIMO functional data. We demonstrate that the PIR Supervector is more powerful than single-variable analysis, and the MIMO PERMANOVA approach outperforms traditional SISO anomaly detection. By distinguishing between normal operation, external thermal disturbance effectively compensated within the investigated operating range, and severe parametric faults that standard univariate methods miss, the framework provides a single, rigorous metric for specific fault diagnosis ($p < 0.001$).}

\onecolumn \maketitle \normalsize \setcounter{footnote}{0} \vfill

\section{\uppercase{Introduction}}
\label{sec:introduction}
Modern chemical processing plants, such as the Tennessee Eastman Process (TEP) \cite{downs1993plant,reinartz2021extended}, a widely used nonlinear and multivariable benchmark for process-control studies, are Multi-Input Multi-Output (MIMO) systems characterized by high dimensionality, non-linearity, and coupling between process variables. In such environments, the dynamic behavior of the plant is not defined by independent trajectories of isolated variables, but by the complex network of functional relationships between manipulated inputs and measured outputs.

Despite this inherent multivariate nature, traditional approaches to process monitoring often rely on isolated analysis. Strategies involve monitoring individual control loops in isolation or reducing dynamic time-series data to static scalar metrics before applying statistical quality control limits \cite{chiang2000fault}. While effective for detecting gross violations of steady-state setpoints, univariate methods are insensitive to systemic changes in the plant's control strategy, shifts in the underlying dynamic transfer functions that may precede failure but do not immediately manifest as scalar outliers \cite{qin2012survey}, unlike continuous coefficient estimation approaches designed to detect slight behavioral changes \cite{icinco18}. 

To overcome the limitations of these simplified representations, this paper introduces a shift in how diagnostic features are extracted and evaluated. The reliance on arbitrary scalar metrics discards the temporal and spectral information embedded within process data. In contrast, describing the system via Frequency Response Functions (FRFs) captures the complete dynamic signature, encompassing gain, phase, and resonance characteristics across varying operational frequencies. A contribution of this work is demonstrating that this functional FRF-based representation provides a superior diagnostic basis compared to traditional scalars, allowing the isolation of subtle parametric faults that manifest as dynamic structural shifts.

\subsection{Motivation: The Multiple Comparisons Problem and Masked Variance}
\label{sec:motivation}
Feedback controllers maintain setpoints, masking the variance of severe faults within individual controlled variables. Because the controller's primary function is to continuously counteract deviations, the underlying fault signature is dynamically transferred from the measured output into the control effort and its correlated variables. Consequently, an anomaly is rarely isolated in a single variable's amplitude; it is contained within the multivariate correlation structure. While multivariate statistical process control (MSPC) tools utilizing Principal Component Analysis (PCA), Partial Least Squares (PLS), or Canonical Variate Analysis (CVA) are more appropriate for managing alarms, as demonstrated in complex systems like cement rotary kilns \cite{Kaced2021}, standard static formulations assume linearity, while non-linear extensions like Kernel PCA assume specific distributional properties unsuitable for complex spectral data \cite{anderson2001new} and fail to capture continuous temporal shifts in underlying dynamic transfer functions. Alternative deep learning approaches, such as automated Convolutional Neural Networks for fault classification \cite{icaart24} or image-based transfer learning for acoustic anomalies \cite{icaart21}, typically demand massive labeled datasets that may be computationally prohibitive for real-time control loops.

Validating control strategies or diagnosing nuanced faults requires comparing plant behavior across multiple operating conditions. This introduces the ``multiple comparisons problem'' \cite{benjamini1995controlling}. When engineers perform repeated pairwise statistical tests across dozens of input-output pairs to characterize a single fault, the probability of obtaining a false positive inflates, rendering standard p-values unreliable. Conversely, applying rigid corrections (e.g., Bonferroni) results in an unacceptable loss of statistical power \cite{peres1999how}.

Frequency Response Functions (FRFs) offer a description of system dynamics \cite{ljung1999system}, analogous to the diagnostic utility of spatial power spectra in robotic environmental analysis \cite{icinco14}, yet remain underutilized in statistical hypothesis testing. This paper proposes a framework bridging this gap. We transform identified FRFs into \textit{Pseudo-Impulse Responses} (PIRs), a robust time-domain representation. By treating these PIRs as discrete functional units within a Permutational Multivariate Analysis of Variance (PERMANOVA) design, we provide a non-parametric statistical test capable of detecting global shifts in MIMO systems. By leveraging permutations rather than assuming multivariate normality, this approach accommodates the highly correlated, high-dimensional nature of functional PIR data without sacrificing statistical rigor.

\subsection{From Anomaly Detection to Strategy Diagnosis}
Real-world process control often asks: \textit{``Is the current process behavior statistically distinct from a specific known state?''} Unlike unsupervised clustering, PERMANOVA requires \textit{a priori} defined groups arising in three scenarios. This supervised grouping allows the framework to explicitly compare the variance between distinct operational states against the natural baseline variability within those states:
\begin{itemize}
\item \textit{Control strategy validation:} When verifying if plant dynamics changed post-modification. The method establishes whether two configurations produce statistically different dynamic responses, though complementary engineering metrics are required to determine which performs better.
\item \textit{Fault diagnosis via library matching:} Testing an unknown state against a library of known failure modes.
\item \textit{Batch process quality assurance:} Testing a current batch against a ``Golden Batch'' reference.
\end{itemize}

This offline, batch-level hypothesis test differs from continuous, online anomaly detection models. The PIR-PERMANOVA framework is specifically designed for rigorous statistical validation of dynamic shifts.

\section{\uppercase{Methods}}
\subsection{The Frequency Response Function (FRF)}
The Frequency Response represents a full description of the steady-state response to a sinusoidal input of frequency $f$. In general, analyzed systems are not strictly linear; the FRF is intended as a representation of an experimental trial \cite{dimaio2024novel}. The FRF $H_{m,i,k}$ for input-output pair index $m$ of subject $i$ at frequency index $k$ is computed using the $H_1$ estimator of input-output pairs excited by a \textit{Pseudo-Random Binary Sequence} (PRBS). This specific estimator is chosen for its robustness against measurement noise present at the system's outputs, yielding a reliable frequency representation even in highly perturbed environments.

\subsection{The Pseudo Impulse Response (PIR)}
For a MIMO system with $N_{in}$ inputs and $N_{out}$ outputs across $N$ experimental runs, we define $M = N_{in} \times N_{out}$ distinct pairs. Let $m$ denote the specific input-output pair index. The PIR for the $i$-th subject for pair $m$ at discrete time $t_j$ is $x_{i,m}(t_j)$, calculated via the transformation \cite{lippi2025bootstrap}:
\begin{equation}
\label{pseudo}
\begin{split}
x_{i,m}(t_j) = 2 \sum^{K}_{k=1} \Big( &\Re(H_{m,i,k}) \cos(2\pi f_{k} t_j) \\
&- \Im(H_{m,i,k})\sin(2\pi f_{k} t_j) \Big).
\end{split}
\end{equation}
Mathematically, this transformation translates complex-valued spectral components back into a tangible time-domain trajectory, preserving phase shifts that scalar metrics discard. The PIR captures the dynamic relationship in the real domain, allowing the application of statistical tools designed for real-valued functions.

\subsection{Extension to MIMO Systems: The PIR Supervector}
To aggregate relevant control loops without biasing the distance metric across vastly different physical quantities, we apply pair-wise z-score standardization. The standardized PIR, $z_{i,m}(t_j)$, represents the dimensionless deviation of subject $i$ from the mean dynamic behavior of pair $m$:
\begin{equation}
z_{i,m}(t_j) = \frac{x_{i,m}(t_j) - \bar{x}_m(t_j)}{\hat{\sigma}_m(t_j)}.
\end{equation}
By enforcing this zero-mean, unit-variance scaling, we ensure that variables operating on fundamentally different physical scales, such as reactor pressures in kilopascals versus valve positions in percentages, contribute equitably to the final statistical evaluation.

The \textit{PIR Supervector} $\mathbf{S}_i$ for subject $i$ concatenates the standardized PIRs of all $M$ pairs into a single vector of length $L = M \times T$:
\begin{equation}
\mathbf{S}_i = \left[ z_{i,1}(t_1), \dots, z_{i,1}(t_T), \ \dots, \ z_{i,M}(t_1), \dots, z_{i,M}(t_T) \right]^\top.
\end{equation}

The frequency grid $\{f_k\}$ was linearly spaced ($0.001$ Hz to $0.5$ Hz), excluding the DC component. The time grid $\{t_j\}$ matched the $10$-hour identification experiment at $1$ Hz. For PERMANOVA, we utilized the Euclidean distance metric on these standardized supervectors ($d_{ij}^2 = \sum (S_{i,k} - S_{j,k})^2$), which captures phase shifts and amplitude variations in linearized dynamics. 

\subsection{The Tennessee Eastman Model (COSTEP)}
We utilized the \textit{COSTEP} Simulink model \cite{VOSLOO2025102217} for the TEP to extract high-resolution telemetry, aligning with the necessity for standardized, multi-task benchmarks when objectively evaluating anomaly detection performance \cite{icissp22}.

\subsection{Experimental Design and Simulated Conditions}
Because the TEP is open-loop unstable, we simulate a standard \textit{Closed-Loop System Identification}. A low-amplitude PRBS ($\pm 0.2$) was superimposed directly onto the Purge Gas Setpoint. Unlike temperature-based excitations which risk thermal runaway (an uncontrollable, self-accelerating positive feedback loop of heat production) in the highly exothermic TEP reactor, perturbing the purge cascade forces closed-loop controllers to actuate the purge valve without violating interlock safety limits. This approach is essential for our framework, as it captures the closed-loop dynamics where the controller's masking behavior is actively engaged. We recorded Reactor Pressure, Reactor Cooling Water Valve Position, and Product G Composition.

Three conditions ($N=10$ runs each) were simulated:
\begin{enumerate}
\item \textit{Condition A (NOC):} Normal Operating Reference.
\item \textit{Condition B (Disturbance 1):} Loss of A feed (IDV 6).
\item \textit{Condition C (Disturbance 2):} Step change in reactor cooling water inlet temperature (IDV 4).
\end{enumerate}

\textbf{Rationale for Scenario Selection:} The two disturbances were selected as physically contrasting test cases to evaluate complementary framework properties. Disturbance 1 (IDV 6) represents a severe alteration of the exothermic mass balance. Although closed-loop controllers may maintain selected variables near setpoints, they do so through substantial changes in coupled input-output dynamics, providing a limiting test for masked faults. Conversely, Disturbance 2 (IDV 4) is an external thermal disturbance that the cooling control loop effectively compensates within the investigated operating range via valve position changes without structurally altering underlying plant dynamics, providing a test of diagnostic specificity.

\subsection{Benchmark Comparison Configuration}
To benchmark against industry standards, static PCA and Dynamic PCA (DPCA) models were constructed, trained exclusively on the reference NOC data. Variables were z-score standardized. The DPCA matrix was augmented with a 5-second lag. Fault detection utilized Hotelling's $T^2$ (measuring Mahalanobis distance in the principal subspace, $T^2 = \mathbf{x}^T \mathbf{P} \boldsymbol{\Lambda}^{-1} \mathbf{P}^T \mathbf{x}$, where $\mathbf{x}$ is the observation vector, $\mathbf{P}$ the loading matrix, and $\boldsymbol{\Lambda}$ the diagonal matrix of retained eigenvalues) and the $Q$-statistic (unmodeled residual error, $Q = \mathbf{e}^T \mathbf{e}$, where $\mathbf{e}$ is the residual vector), averaged over time samples to yield a single scalar metric per run. In physical terms, $T^2$ monitors variations within the known acceptable operating bounds, whereas $Q$ serves as a critical indicator for novel events that entirely rupture the previously modeled multivariate correlations.

\section{\uppercase{Results}}
\label{sec:results}

\subsection{The Failure of Univariate Anomaly Detection}
\label{sec:siso_example}
\begin{figure*}[ht]
\centering
\includegraphics[clip, trim=2.5cm 0.5cm 2.5cm 0.5cm,width=1\textwidth]{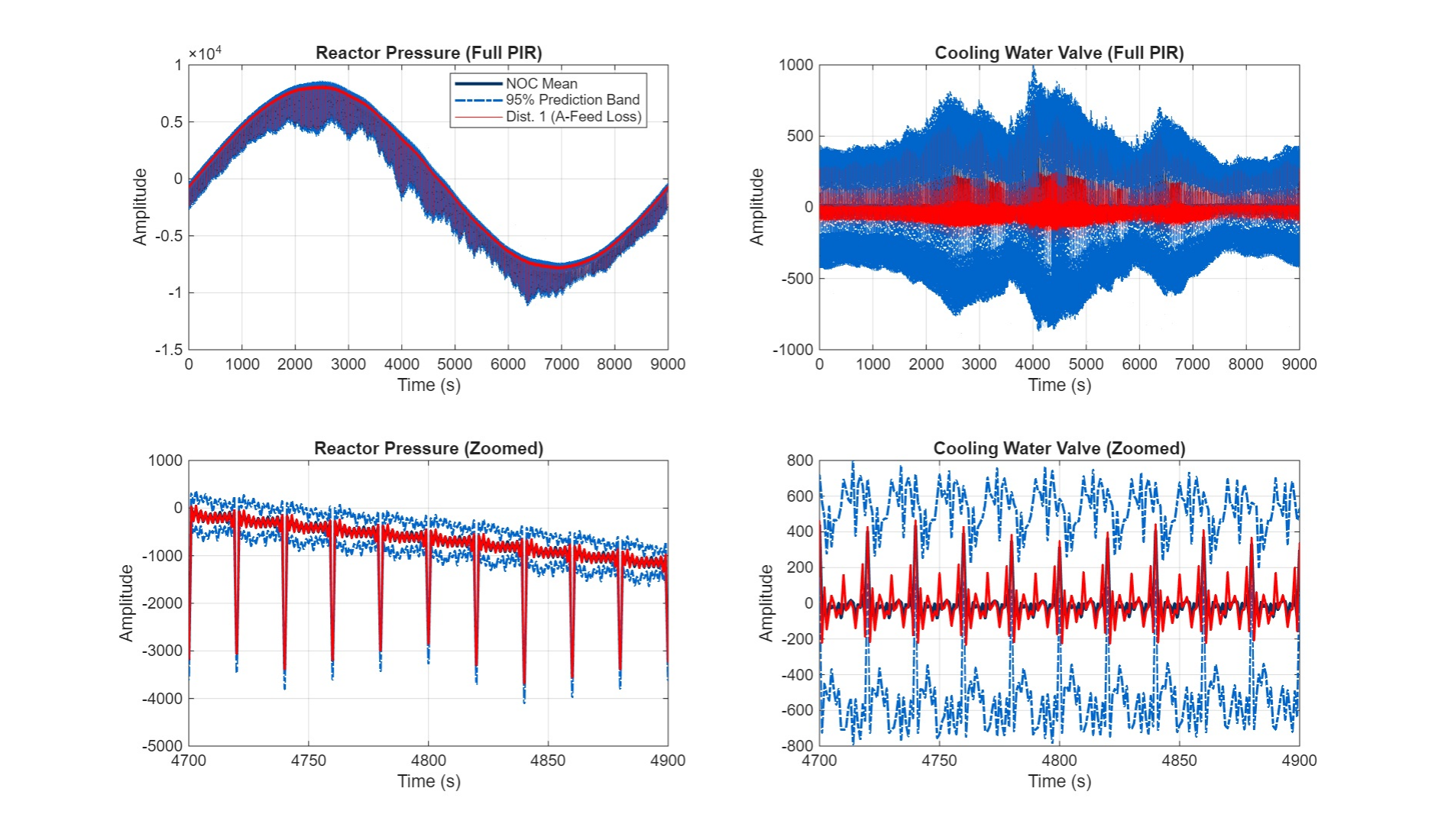} 
\caption{\textit{The Failure of Univariate (SISO) Anomaly Detection.} Time-domain Pseudo-Impulse Responses (PIR) for Reactor Pressure (Left) and Cooling Water Valve (Right). The blue shaded regions dictate the 95\% bootstrap prediction bands established under Normal Operating Conditions (NOC). The red line represents a severe A-Feed Loss fault (Disturbance 1). Due to realistic process noise and aggressive closed-loop control, the univariate fault trajectory remains entirely constrained within the healthy 95\% prediction bands, resulting in a false negative.}
\label{fig:siso_example}
\end{figure*}

Before evaluating the MIMO framework, we establish why univariate anomaly detection fails. Utilizing bootstrap prediction bands, we established a 95\% reference condition (NOC) for individual loops. When the plant experiences a severe A-feed loss (Disturbance 1), the plant-wide decentralized control configuration described by \cite{chiang2000fault}, hereafter referred to as the Braatz controller, maintains variables close to setpoints. As illustrated in Figure \ref{fig:siso_example}, this masks the fault's variance, yielding a false negative. Because the bootstrap bands evaluate the amplitude of each variable independently, they are blind to the structural decoupling occurring across the system. The anomaly is hidden within the multivariate correlation structure.

\subsection{Global Test of Differences}
Unlike the SISO approach, the global PERMANOVA test evaluates the holistic correlation structure. The test yielded a significant separation ($N=10$ per group, $9999$ permutations):
\begin{itemize}
    \item \textbf{Explained Variance ($R^2$):} $0.1650$ (16.5\%)
    \item \textbf{Residual Variance:} $0.8350$ (83.5\%)
    \item \textbf{Pseudo-F Statistic:} $2.6679$ \ \textbf{p-value:} $0.0003$
\end{itemize}

A homogeneity of multivariate dispersions test (PERMDISP) yielded $p = 0.8574$, confirming dispersions are homogeneous. This is a prerequisite; it suggests that the detected anomaly reflects a translation of the system's dynamic centroid rather than an inflation of operating noise. Consequently, the significant PERMANOVA result is driven by true spatial centroid shifts (the parametric faults). The observed effect size (Cohen's $f^2 = 0.1976$) indicates a moderate separation. However, this result should not be interpreted as a formal prospective power analysis, and the limited number of independent runs is acknowledged as a limitation of this proof-of-concept study. 

The 83.5\% Residual Variance represents the process noise combined with the dynamic masking actions of the controller. The 16.5\% Explained Variance ($R^2$) isolates the systemic shifts caused explicitly by the faults. The supervector separates this signal from the masking effects, yielding $p=0.0003$. In essence, the permutation test acts as a mathematical sieve, separating the structured, fault-induced variance from the aggressive stochastic masking of the feedback loop.

\begin{figure}[ht]
\centering
\includegraphics[width=0.9\columnwidth]{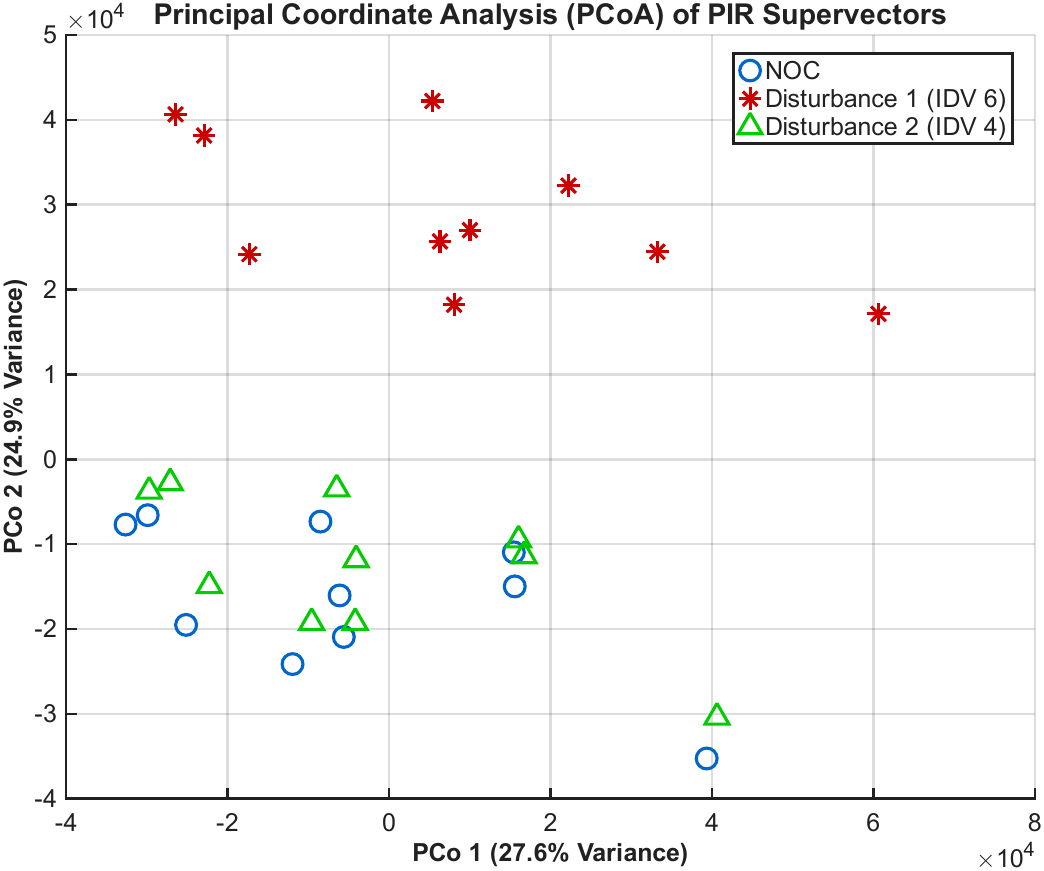} 
\caption{\textit{Principal Coordinate Analysis (PCoA) of PIR Supervectors.} The scatter plot represents the first two principal coordinates derived from the squared Euclidean distance matrix. The distinct spatial separation of Disturbance 1 (IDV 6) visually demonstrates the systemic dynamic shift detected by the PERMANOVA ($p < 0.0001$), whereas Disturbance 2 (IDV 4) heavily overlaps with the normal operating baseline, preventing false alarms.}
\label{fig:pcoa}
\end{figure}

To visually demonstrate the spatial centroid shifts confirmed by the PERMANOVA and PERMDISP tests, a Principal Coordinate Analysis (PCoA) was conducted. The squared Euclidean distance matrix was extracted from the standardized PIR Supervectors, and Classical Multidimensional Scaling was applied to map the multivariate variance into a two-dimensional subspace. As illustrated in Figure \ref{fig:pcoa}, the severe parametric failure (Disturbance 1) forms a distinct, separated cluster from the Normal Operating Conditions. Conversely, the environmental shift (Disturbance 2) remains tightly integrated with the NOC baseline, reinforcing the specificity of the proposed framework.

\subsection{Pairwise Comparisons}
To isolate the catastrophic parametric failure from the environmental shift, post-hoc pairwise comparisons were conducted (Table \ref{tab:pairwise}). By applying a Bonferroni correction, we neutralize the multiple comparisons problem highlighted earlier, ensuring that any flagged difference represents a physical alteration.

\begin{table}[h]
\caption{Pairwise PERMANOVA comparisons between operating conditions. P-values are adjusted using the Bonferroni correction ($p_{adj} = p_{raw} \times 3$).}
\label{tab:pairwise}
\centering
\small 
\setlength{\tabcolsep}{4pt} 
\begin{tabular}{|l|c|c|}
\hline
\textit{Comparison} & \textit{F} & \textit{$p_{adj}$} \\ \hline
NOC vs. Dist. 1 & 3.8554 & \textit{$<$ 0.0001} \\ \hline
NOC vs. Dist. 2 & 0.1019 & \textit{1.0000} \\ \hline
Dist. 1 vs. Dist. 2 & 3.6034 & \textit{$<$ 0.0001} \\ \hline
\end{tabular}
\end{table}

\section{\uppercase{Discussion}}
\label{sec:discussion}

\subsection{Specificity and Robustness to Additive Faults}
When subjected to Disturbance 2 (IDV 4), the PERMANOVA yielded $p = 1.0000$. Because this is an additive disturbance, the Braatz cascade controller compensates by opening the cooling valve to a new steady-state position. This compensation shifts the operating point without distorting the physical coupling between inputs and outputs; therefore, the thermodynamic transfer function is not significantly different. The PIR mathematically rejected the disturbance, avoiding a false alarm.

\subsection{Sensitivity to Parametric Failures}
Conversely, Disturbance 1 (IDV 6) was detected with statistical significance ($p < 0.0001$). A feed loss significantly alters the mass balance. Unlike a simple additive offset, this parametric failure alters the dynamic coupling of the entire system, creating a persistent structural change that isolated sensors cannot contextualize. The PIR supervector aggregates the masked variables (Figure \ref{fig:siso_example}) with corresponding reactive shifts, providing a single p-value for the entire system state without inflating the Family-Wise Error Rate, sharing the conceptual objective of fault training matrices without discretizing the continuous dynamic profile \cite{icinco19}.

\subsection{Direct Comparison with Static and Dynamic PCA}

\begin{figure*}[ht]
\centering
\includegraphics[width=1\textwidth]{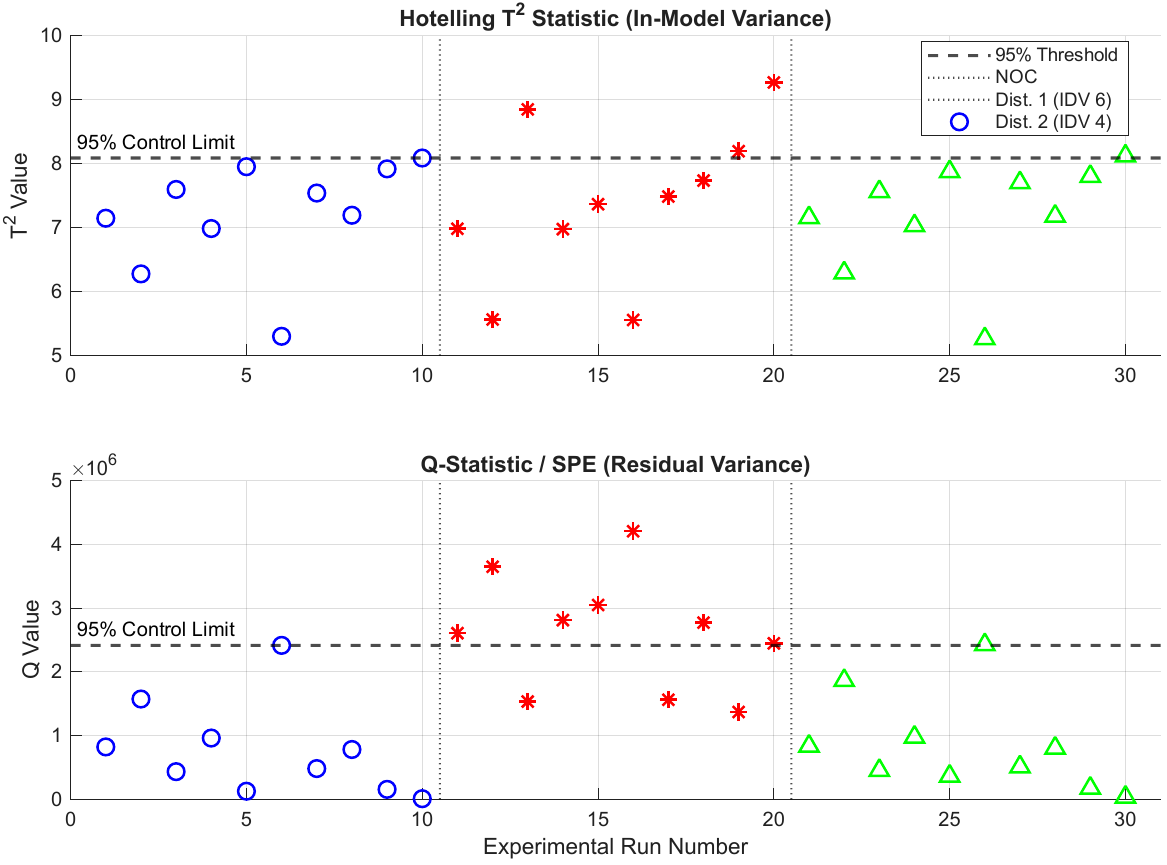} 
\caption{\textit{Baseline PCA performance on the MIMO TEP dataset.} Each marker represents one independent simulation run. Blue circles indicate NOC (runs 1–10), red asterisks indicate Disturbance 1 (runs 11–20), and green triangles indicate Disturbance 2 (runs 21–30). Vertical dotted lines separate the three operating conditions, while the horizontal dashed line represents the 95\% control limit. \textbf{Top:} Hotelling's $T^2$ statistic. \textbf{Bottom:} The $Q$-statistic.}
\label{fig:pca_baseline}
\end{figure*}

As illustrated in Figure \ref{fig:pca_baseline}, standard projection methods are limited when confronted with parametric dynamic faults. The baseline PCA and DPCA models required 8 and 7 components respectively to explain 95\% of the NOC variance. The $Q$-statistic shows an elevation for the severe A-Feed Loss, acting as a binary alarm. However, Hotelling's $T^2$ demonstrates significant overlap across conditions. DPCA evaluates multivariate data as discrete, localized snapshot matrices. By treating time as a series of isolated frames rather than a continuous functional trajectory, these projection methods discard the transient phase relationships for diagnosing structural shifts. Consequently, severe dynamic faults and environmental shifts both manifest as overlapping clusters that struggle to breach the 95\% limit consistently. By linearizing continuous frequency responses into standardized time-domain vectors, the PIR framework natively captures holistic temporal inertia, yielding statistically significant separations ($p < 0.0001$) while avoiding the complex neural network cascades required by some nonlinear PCA methods \cite{ncta12}.

\subsection{Methodological Limitations and Framework Boundaries}
\label{sec:limitations}
While this proof-of-concept demonstrates the efficacy of the PIR-PERMANOVA framework, several methodological boundaries must be explicitly defined. First, the experimental implementation represents a SIMO (Single-Input, Multi-Output) special case of the generalized MIMO formulation. Specifically, the PRBS excitation was applied to a single input ($N_{in}=1$) while monitoring three outputs ($N_{out}=3$), yielding $M=3$ input-output pairs. With a $10$-hour experiment sampled at $1$ Hz ($T=36000$ samples), the resulting supervector length is $L=108000$. The summation in Equation \ref{pseudo} runs over $K=18000$ frequency lines to perfectly reconstruct the time grid, with the factor of two accounting for the complex conjugate pairs of the real inverse discrete Fourier transform convention. 

Second, applying an FRF, a fundamentally linear system descriptor, to the inherently nonlinear TEP yields a linearized operational approximation. While the $H_1$ estimator extracts the optimal linear fit under the specified PRBS excitation, assertions that this captures the ``complete'' dynamic signature are strictly bound to this linearized subspace. Highly non-linear transient behaviors that cannot be approximated by phase and gain shifts may remain undetected.

Third, combining Hotelling's $T^2$ and the $Q$-statistic via a simple temporal average within each run for the benchmark comparison merges metrics with fundamentally different scales and physical interpretations (in-subspace variation versus residual error). While this mirrors standard dimensionality reduction practices to yield a singular diagnostic scalar, it inherently obscures the specific failure modality that DPCA is designed to track.

Finally, the statistical conclusions are bound by the sample size ($N=10$ per group). The lack of statistical significance in the PERMDISP test ($p = 0.8574$) fails to detect dispersion differences but does not deterministically guarantee their absence due to the low statistical power inherent to small samples. Consequently, the ``more powerful'' characterization of the PIR framework reflects its empirical sensitivity to structural faults relative to the univariate baselines, rather than the conclusion of a formal prospective power analysis. Future work will benchmark the framework across a massively scaled simulation matrix to formally establish its Type II error limits.

\subsection{Conclusions}
\label{sec:conclusions}
The \textit{PIR supervector} distinguishes between Normal Operating Conditions and distinct disturbance scenarios with statistical confidence ($p < 0.001$). By penetrating the stochastic masking of closed-loop controllers, this functional approach provides operators with a tool to differentiate between environmental drift and critical system degradation. The present results should not be interpreted as an exhaustive validation across all TEP failure modes. Such an assessment constitutes an important direction for future work. Future work will also focus on leveraging spatial structures to generate \textit{Strategy Heatmaps}, incorporating Principal Coordinate Analysis (PCoA) visualizations to map the distance matrices, and applying expert-defined weightings to construct a \textit{safety super vector} prioritizing critical control loops based on specific industrial risk profiles. Furthermore, while the PIR-PERMANOVA framework is applied here to the Tennessee Eastman Process for the first time, the foundational methodology of evaluating FRFs via PIRs was originally pioneered for human posture control. Consequently, future investigations will adapt this multiple-group comparison strategy to solve complex multi-class diagnostic scenarios in human movement and sensorimotor control \cite{lippi2025head,LIPPI2023139,icinco23}.

\section*{\uppercase{Acknowledgements}}
The library implementing FRF tests and specifically the new PERMANOVA test is available at: \small {https://github.com/VittorioFreiburg/FRF-statistics} \cite{lippi2024frf,lippi2026}. This work is supported by Calejo Hybrid Intelligence AB. \\
\vspace{-0.6cm}
\bibliographystyle{apalike-ejor}
{\small
\bibliography{example}}

\end{document}